\documentclass[aps,prd,11pt]{revtex4}
\usepackage[colorlinks=true, pdfstartview=FitV, linkcolor=blue, citecolor=red, urlcolor=magenta]{hyperref}
\usepackage{graphicx}
\usepackage{latexsym}
\usepackage{amsmath}
\usepackage{amsfonts}
\usepackage{amssymb}
\usepackage{verbatim}

\newcommand{\be}{\begin{equation}}
\newcommand{\ee}{\end{equation}}
\newcommand{\bea}{\begin{eqnarray}}
\newcommand{\eea}{\end{eqnarray}}

\newcommand{\ben}{\begin{eqnarray}}
\newcommand{\een}{\end{eqnarray}}

\begin{document}

\title{Holographic description of confinement and screening through brane cosmology}
\author{F.A. Brito$^{1,2}$, D.C. Moreira$^{1,2}$ } 
\affiliation{$^{1}$Departamento de F\'\i sica,
Universidade Federal de Campina Grande, Caixa Postal 10071,
58109-970  Campina Grande, Para\'\i ba,
Brazil \\
$^{2}$Departamento de F\'\i sica,
Universidade Federal da Para\'\i ba, Caixa Postal 5008,
58051-970 Jo\~ ao Pessoa, Para\'\i ba,
Brazil
}

\begin{abstract}
We compute the holographic quark potential in the realm of brane cosmology. We show that under certain conditions the very geometry due to an inflationary 3-brane induces a D3-D7-brane system.  
The cosmological constant that appears involved in the original geometry is attributed to the D7-brane position itself in its embedding process.
We address the issues of confinement at low distances, screening effects at sufficiently large distances, and quark condensate. 
 
\end{abstract}
\pacs{11.15.-q, 11.10.Kk} \maketitle


\section{Introduction}

In order to extend the AdS/CFT correspondence \cite{Maldacena:1998im} to accomplish QCD-like theories one needs to break supersymmetry and conformal invariance. The AdS/CFT correspondence is formulated around the background geometry of a stack of $N$ D3-branes where the strings may end --- for a review see \cite{D'Hoker:2002aw, myers, Erdmenger:2007cm}. The corresponding field theory is a gauge theory whose fields are in the adjoint representation of $SU(N)$ group. To take into account flavor degrees of freedom one needs to add $N_f$ flavor branes. Now strings can be stretched between different types of branes with only one charge under the $SU(N)$
group on the D3-branes and then describe quark fields. The corresponding field theory now has fields (quarks) that are in the fundamental representation. One of the most interesting set up to accomplish this is the D3-D7-brane system with $N_f$ D7 flavor branes in the probe limit ($N_f\ll N$) in order not to  change the background geometry \cite{Karch:2002sh,Karch:2002xe}. Strings can have both ends on the flavor brane to describe a dual to quark-antiquark operators since they are
in the adjoint representation of $SU(N_f)$ \cite{Erdmenger:2007cm}. Such effort is one among several others in the direction to achieve QCD-like theories.

In a general perspective one can think of this proposal as a way to `deform' the $AdS$ space. For instance, for confining theories see  \cite{BoschiFilho:2004ci,BoschiFilho:2005mw,Csaki:2008dt,Klebanov:2000hb,Maldacena:2000yy,BoschiFilho:2006pe,Edelstein:2006kw}. In  \cite{bbq08} was assumed a deviation from the conformal case which
is {\it naturally} obtained in geometries of brane cosmology scenarios \cite{Binetruy:1999ut, Binetruy:1999hy}. As we shall see, a small deviation around AdS space controlled by the tension and the cosmological constant on the brane is encoded on a fundamental constant $C$. It was found that such a
deviation that leads to a {\it confining regime} in the dual gauge field theory is directly related to cosmological constant on the brane. Indeed, this encloses a general class of geometries where the brane has a
cosmological constant, i.e., a {\it bent} brane, such as $AdS_4$ or $dS_4$ branes \cite{Cvetic:1993xe,Kaloper:1999sm,Karch:2000ct,Bazeia:2006ef,Bazeia:2007vx,Bazeia:2004yw,Ghoroku:2006nh,Erdmenger:2011sy},
in contrast with flat D3-branes in the original AdS/CFT correspondence. Other inflationary brane scenarios can also be found in \cite{Bazeia:2007vx,Dvali:1998pa,Lukas:1998yy,Chamblin:1999ya,Lukas:1999yn}.

In this paper we revisit the previous results in \cite{bbq08} to address the regimes of confinement and screening from the perspective of D3-D7-brane system. As we shall show, by making suitable transformation of coordinate and embeddings the very geometry due to an inflationary 3-brane \cite{Binetruy:1999ut, Binetruy:1999hy} induces a D3-D7-brane system, in Euclidean signature. For further details on Euclidean brane cosmology see \cite{Savonije:2001nd,Medved:2001jq}. The D7-brane position will be directly related to the constant $C$.  In addition to the several reasons raised above to choose such a system, we recall that from the point of view of cosmological selection of the number of spatial dimensions, a ten-dimensional universe initially filled with a dilute D$p$-anti-D$p$-brane gas tends to find its equilibrium filled with only D3 and D7-branes in the realm of type IIB superstrings \cite{Karch:2005yz}. 

The paper is organized as follows. In Sec.~\ref{sec1} we explore the emergence of an induced D7-brane through brane cosmology. In Sec.~\ref{sec2} we address the issue of the D7-brane embedding and show that the quark condensate is a linear function of the quark mass. It should be stressed that the flavoring is due to the presence of the cosmological constant.
In Sec.~\ref{sec3}, we compute the interquark potentials from the holographic (`AdS/QCD') point of view \cite{Maldacena:1998im,Rey:1998ik,Minahan:1998xq,Greensite:1999jw,Witten:1998zw,Polyakov:2000ti,Bigazzi:2004ze,Sonnenschein:2000qm,BoschiFilho:2004ci,BoschiFilho:2005mw,Andreev:2006ct,Csaki:2008dt,Drukker:1999zq} by relating the string world-sheet area along the bulk with the area of the Wilson loop on the boundary in Euclidean signature. Finally, in Sec.~\ref{conclu} we present our final discussions.

\section{Brane cosmology and the emergence of induced D7-brane}
\label{sec1}

We shall mainly focus on the fact that the brane cosmology metric \cite{Binetruy:1999ut,Binetruy:1999hy} (in its Euclidean form) as has been shown in \cite{bbq08} can be written as follows
\bea\label{ads5}
ds_5^2=\alpha'\left[\frac{U_0^2}{R^2}y^2\left(dt^2+a_0^2\gamma_{ij}dx^idx^j\right)+\frac{R^2}{y^2-C}dy^2\right],
\een
where
\ben\label{constant-C}
C=\frac{2\gamma R^2}{\alpha' U_0^2}, \qquad \gamma=3H_E^2/2\sigma, \qquad \gamma_{ij}=\delta_{ij}, \qquad i,j=1,2,3
\een
being $H_E$ the Euclidean Hubble parameter defined in terms of an induced Friedmann equation on the inflating 3-brane with tension $\sigma$ \cite{Binetruy:1999ut,Binetruy:1999hy, bbq08} that we identify with the D3-brane tension $\tau_{3}$ \cite{bbq08}. We also assume flat universe ($k=0$) thus $\gamma_{ij}=\delta_{ij}$. Notice that the coordinate $U=U_0y$ is assumed to have dimension of energy, such that $U_0^2\to U_0^2/\alpha'$ and $R^4\equiv4\pi g_sN\alpha'^2\to R^4\equiv4\pi g_sN$. We should note that this metric can be rewritten in a suitable form by {\it changing the variables} $\rho^2=U_0^2y^2-\tilde{C}$ and $\tilde{C}=U_0^2C$ such that we find
\bea\label{ds5}
ds_5^2=\alpha'\left[\frac{\rho^2+\tilde{C}}{R^2}\left(dt^2+a_0^2\gamma_{ij}dx^idx^j\right)+\frac{R^2}{\rho^2+\tilde{C}}d\rho^2\right]
\een
This metric can be seen as the induced metric on a D7-brane if we want our setup to be developed by a geometry in a ten-dimensional spacetime generated by  a stack of $N$ D3-brane and a number of probe D7-branes ($N_f$ D7 flavor branes) \cite{Karch:2002sh,Karch:2002xe} where the strings can end. Since we are considering probe D7-branes ($N_f\ll N$), there is a way to turn the original metric to an induced D7-brane metric by keeping it as a solution of the original equations of motion. Thus, let us first complete the metric (\ref{ds5}) by extending the number of dimensions involved in the `radial' coordinate as follows 
\bea
\rho^2=X_4^2+X_5^2+X_6^2+X_7^2\qquad d\rho^2\to d\rho^2+\rho^2d{\Omega_3^2}=dX_4^2+dX_5^2+dX_6^2+dX_7^2
\eea
such that we have the induced metric on a D7-brane
\bea\label{ds8}
ds_8^2=\alpha'\left[\frac{\rho^2+L^2}{R^2}\left(dt^2+a_0^2\gamma_{ij}dx^idx^j\right)+\frac{R^2}{\rho^2+L^2}\left(d\rho^2+\rho^2d{\Omega_3^2}\right)\right]
\een
where $L\equiv\sqrt{\tilde{C}}=U_0\sqrt{C}$. One should notice that for $\rho\to 0$ (which means that $U_0^2y^2\sim L^2$, i.e., near the place where the D7 brane sits in the 89-plane --- see below) the radius of $S^3$ shrinks to zero and the $ds_8^2$ metric  approaches the original $ds_5^2$ metric.
As we shall see later, $L$ is the place where the D7-brane sits and will be directly related to the quark mass $m_q$. Notice that for $\rho\gg L$ (or for $L=0$) the metric approaches a geometry with the topology $AdS_5\times S_3$ which simple reflects the absence of dynamical quarks with light masses $m_q$ and the strings turn out to be ending on the $AdS$ boundary generating quarks with infinite mass. In a general situation the strings end on the D7-branes generating quarks with light masses presenting both confinement and flavoring. 

\section{Flavoring with the presence of the cosmological constant}
\label{sec2}

The embedding of the D7-brane. To address the issue of D7-brane embedding one should consider the complete metric in a ten-dimensional spacetime
\bea\label{eq-10d}
ds_{10}^2=\alpha'\left[\frac{\rho^2+X_8^2+X_9^2}{R^2}\left(dt^2+a_0^2\gamma_{ij}dx^idx^j\right)+\frac{R^2}{\rho^2+X_8^2+X_9^2}\left(d\rho^2+\rho^2d{\Omega_3^2+dX_8^2+dX_9^2}\right)\right]
\eea
that by using symmetry in the $X_8-X_9$ plane and setting $X_9=0$ and $X_8=\omega(\rho)$ we find the induced metric for the D7-brane
\bea\label{embed-induced}
ds_{8}^2=\alpha'\left[\frac{\rho^2+\omega(\rho)^2}{R^2}\left(dt^2+a_0^2\gamma_{ij}dx^idx^j\right)+\frac{R^2}{\rho^2+\omega(\rho)^2}\left((1+(\partial_\rho\omega)^2)d\rho^2+\rho^2d{\Omega_3^2}\right)\right]
\eea
The embeddings $\omega(\rho)$ is obtained by solving the equation of motion deduced from the D7-brane action given up to angular factors by 
\ben
S_{D7}=-\tau_7\int{d^8\xi\,a_0^3\,\rho^3\sqrt{1+\omega'(\rho)^2}},
\een
where $\omega'(\rho)\equiv\partial_\rho\omega$ and the D7-brane tension $\tau_7=[(2\pi)^7g_s\alpha'^4]^{-1}$. 
We have used the metric (\ref{embed-induced}) and disregarded the gauge potentials on the D7-brane world-volume. The equation of motion
\bea\label{d7-eom-0}
\frac{d}{d\rho}\left[\frac{\rho^3}{\sqrt{1+\omega'(\rho)^2}}\frac{d\omega}{d\rho}\right]=0
\eea
has the well-known asymptotic solution given by
\ben\label{w-asymp}
\omega(\rho)\sim L+\frac{c}{\rho^2}, \qquad \rho\to\infty
\een
where $c$ corresponds to the v.e.v. of an operator that has dimension three since $\rho$ carries energy dimension and measures the
quark condensate $<\bar{\psi}\psi>$. $L$ is an operator with energy dimension previously defined as
\bea
L=U_0\sqrt{C}=\frac{\sqrt{2}\sqrt{\gamma}R}{\sqrt{\alpha'}}
\eea
where we have used the explicit form of $C$ given in (\ref{constant-C}). Now considering $L\equiv 2\pi m_q$ allows us to define the quark mass $m_q$. 
It  can be written explicitly as follows   
\begin{equation}\label{mass-mq}
m_q=\frac{1}{\pi\sqrt{2}}\sqrt{\gamma}\frac{R}{\sqrt{\alpha'}}.
\end{equation}
Here instead of solving numerically Eq.~(\ref{d7-eom-0}) to find the embeddings we take advantage of using the asymptotic relationship between $\omega(\rho)$ and $U$ previously stated.
Thus, now starting with the original cosmological solution \cite{bbq08} we can see that 
\bea
U^2(r)=\gamma+\xi^2e^{\mu r}+\chi^2e^{-\mu r}\to U^2(r)=\xi^2e^{\mu r} \equiv \rho^2 \qquad \mbox{as}\qquad\: r\to\infty
\eea
that solving for $\exp(\mu r)$ we find asymptotically 
\ben
U^2(r)=\gamma+\rho^2+\frac{\xi^2\chi^2}{\rho^2}, \qquad \rho\to\infty
\een
Now restoring the dimension of energy ($U^2\to U^2/\alpha'$, with dimensionless $R$), recognizing that $\omega^2(\rho)=U^2-\rho^2\sim L^2$ asymptotically (which is consistent with change of variables in the metric (\ref{ds5})) and using the fact that $\xi^2\chi^2=\frac{\gamma^2}{4}$, which is restricted to the region $\gamma \leq 1/2$ \cite{bbq08} we have 
\ben\label{w-rho-2}
\omega^2(\rho)=\frac{\gamma}{\alpha'}+\frac{\gamma^2}{4\alpha'^2\rho^2}\to \omega(\rho)\simeq \sqrt{\frac{\gamma}{\alpha'}}\left(1 +\frac{\gamma}{8\alpha'\rho^2} \right)
\een 
Comparing with Eq.~(\ref{mass-mq}) we can rewrite Eq.~(\ref{w-rho-2}) as 
\bea
\omega(\rho)\simeq \frac{2\pi m_q}{\sqrt{2}R}+\frac{(2\pi m_q)^3}{8{(\sqrt{2}R)}^{3}\rho^2}.
\eea
Notice that this formula is dimensionally consistent with the asymptotic solution Eq.~(\ref{w-asymp}). Thus, up to a common $\sqrt{2}R$ factor, we simply find 
\bea\label{w-2-M}
\omega(\rho)&\simeq& {2\pi m_q}+\frac{2\pi^2 m_q^2}{R^2}\frac{2\pi m_q}{8\rho^2}\nonumber\\
&\simeq& {2\pi m_q}+\frac{2\pi m_q M^2}{8\rho^2},\qquad \mbox{where} \qquad M^2=\frac{2\pi^2 m_q^2}{R^2}=\pi\sigma_c
\een
As we shall show later, $\sigma_c$ is the tension of the confining `QCD' string. The condensate operator can be now identified with 
\bea
c=\frac14{\pi m_q M^2}.
\eea
For fixed $\sigma_c$,  the condensate increases {\it linearly} with the quark mass $m_q$ \cite{Chiu:2003iw,Fleming:1998cc,Burger:2012ti}. This seems to be natural for light $m_q$ masses \cite{Erdmenger:2007cm}. Notice that for $M=m_q=m$ this relationship between quark condensate and quark mass has an interesting resemblance to {\it Ioffe formula} \cite{wolfram} for nucleon masses
\bea
m\approx (4\pi^2<\bar{\psi}\psi>)^{1/3}.
\eea

\section{The interquark potential}
\label{sec3}

In order to follow with our objective such as to obtain the energy and inter-quark distances we take the relevant part of the $AdS_5$ metric of the background (\ref{eq-10d}) 
\bea\label{ads5-2}
ds_{5}^2=\alpha'\left[\frac{X_4^2}{R^2}\left(dt^2+a_0^2\gamma_{ij}dx^idx^j\right)+\frac{R^2}{X_4^2}dX_4^2\right]
\to\alpha'\left[\frac{U_0^2}{R^2}y^2\left(dt^2+a_0^2\delta_{ij}dx^idx^j\right)+\frac{R^2}{y^2}dy^2\right],
\een
This geometry can be identified with a near horizon geometry of a stack of $N$ D3-branes located at $X_4=X_5=...=X_9=0$ and should not be modified by the $N_f$ D7-branes as long as one assumes the probe limit $N_f\ll N$.

The distance  between quarks is given by \cite{Sonnenschein:2000qm}
\begin{equation}
\tilde{L}=2\int^{r1}_{r_0}dr\,\frac{g(\tau,r)}{f(\tau,r)}\frac{f(\tau,r_0)}{\sqrt{f^2(\tau,r)-f^2(\tau, r_0)}}=\frac{L_0}{{a}_0(t)}
\end{equation}
where using the original metric given in brane cosmology \cite{Binetruy:1999ut,Binetruy:1999hy,bbq08} $g^2(\tau,r)=g_{00}g_{rr}=U^2$ and $f^2(\tau,r)=g_{00}g_{xx}= U^4 {a}_0^2$
we find
\begin{equation}
L_0=2\int^{r_1}_{r_0}dr\frac{1}{U}\frac{U_0^2}{\sqrt{U^4-U_0^4}}
\end{equation}
which is in general the `co-moving inter-quark distance'. Finally using the coordinates adopted in the metric (\ref{ads5-2}) we find the useful formula
\begin{equation}
L_0=
2\frac{R^2}{U_0}\int^{U_1/U_0}_{1}\frac{dy}{y^2\sqrt{y^4-1}} \label{lo}
\end{equation}
Similarly we have the Nambu-Goto action \cite{bbq08,Sonnenschein:2000qm}
\begin{eqnarray}
\nonumber S&=&2\frac{1}{2\pi\alpha'}\int^{r_1}_{r_0}\int^T_0 dr d\tau\frac{g(\tau,r)f(\tau,r)}{\sqrt{f^2(\tau,r)-f^2(\tau,r_0)}}
 =2\frac{T}{2\pi\alpha'}\int^{r_1}_{r_0} dr \frac{U^3}{\sqrt{U^4-U_0^4}}.
\end{eqnarray}
Again using the coordinates adopted in the metric (\ref{ads5-2}) and the expectation value of the Wilson loop $\langle W(C) \rangle\sim e^{-E(L_0)T}\sim e^{-S}$ where $S$ is the Nambu-Goto action we find the useful formula for the inter-quark energy
\begin{eqnarray}\label{En}
 E=
 2\frac{ U_0}{2\pi}\int_1^{U_1/U_0}\frac{y^2}{\sqrt{y^4-1}}dy.
\end{eqnarray}
The equations (\ref{lo}) and (\ref{En}) are fundamental to obtain the inter-quark potentials and to study several regimes. 
In both cases we shall assume the endpoints $U_1=L=2\pi m_q=U_0\sqrt{C}$ of the string symmetric profile ending where the D7-brane sits. Thus, we are understanding the upper limit as $U_1/U_0\equiv\sqrt{C}$. In the following we are going to consider 
 large and small separation between the quarks, as compared to the scale of the lightest meson.

\subsection{Large interquark distance}

In this case we should consider $U_1/U_0\equiv\sqrt{C}\gg1$ \cite{myers}. Firstly for $L_0$ we rewrite Eq.~(\ref{lo}) as follows
\begin{eqnarray}
\frac{L_0}{2}=\frac{R^2}{U_0}\int^{U_1/U_0}_{1}\frac{dy}{y^2\sqrt{y^4-1}} &=&\frac{R^2}{U_0}\int^{\infty}_{1}\frac{dy}{y^2\sqrt{y^4-1}}-\frac{R^2}{U_0}\int_{U_1/U_0}^{\infty}\frac{dy}{y^2\sqrt{y^4-1}}\nonumber\\
&\simeq&\frac{R^2}{U_0}\kappa-\frac{R^2}{U_0}\int_{U_1/U_0}^{\infty}\frac{dy}{y^4}\nonumber\\
&\simeq&\frac{R^2}{U_0}\kappa-\frac13\frac{R^2U_0^2}{U_1^3}
\label{lo2}
\end{eqnarray} 
In the above integrations at r.h.s., the first term gets $\kappa=\frac{1}{\sqrt{2\pi}}\Gamma\left(\frac{3}{4}\right)^2$ and the integrand in the second term is well approximated by $y^{-4}$ since $y\gg1$ in that interval. Now recalling that $U_1=2\pi m_q$
we find 
\begin{eqnarray}
\frac{L_0}{2}\simeq\frac{R^2}{U_0}\kappa-\frac13\frac{R^2U_0^2}{(2\pi m_q)^3}
\label{lo2-3}
\end{eqnarray} 
Let us now solve this equation iteratively for $U_0$ to obtain
\begin{eqnarray}
U_0\simeq\frac{2R^2\kappa}{L_0}\left[1-\frac13\frac{R^6\kappa^2}{(\pi L_0m_q)^3}\right]
\label{lo2-4}
\end{eqnarray} 
It is interesting to note that taking the leading term we find
\begin{equation}
\frac{m_q}{U_0}\simeq\frac{m_q L_0}{2R^2\kappa}\sim L_0m_{\rm gap},
\end{equation}
where $m_{\rm gap}\sim m_q/\sqrt{g_sN}$. This implies that our approximation is valid as long as $L_0m_{\rm gap}\to\infty$.

Now for energy we rewrite Eq.~(\ref{En}) as follows
\begin{eqnarray}\label{En2}
 E=\frac{ U_0}{\pi}\int_1^{U_1/U_0}\frac{y^2}{\sqrt{y^4-1}}dy=\frac{ U_0}{\pi}\int_1^{\infty}\frac{y^2}{\sqrt{y^4-1}}dy-\frac{ U_0}{\pi}\int^\infty_{U_1/U_0}\frac{y^2}{\sqrt{y^4-1}}dy,
\end{eqnarray}
that regularizing each integral in the r.h.s. by properly subtracting the diverging terms
\begin{eqnarray}
\int_0^{\infty}{dy}=\int_{1}^{\infty}{dy}+\int_0^{1}{dy}, \qquad \int_0^{\infty}{dy}=\int_{U_1/U_0}^{\infty}{dy}+\int_0^{U_1/U_0}{dy}
\label{reg}
\end{eqnarray} 
we  find 
\begin{eqnarray}\label{En2-reg}
 E=2m_q+\frac{ U_0}{\pi}\left[\int_1^{\infty}\left(\frac{y^2}{\sqrt{y^4-1}}-1\right)dy-1\right]-\frac{ U_0}{\pi}\int^\infty_{U_1/U_0}\left(\frac{y^2}{\sqrt{y^4-1}}-1\right)dy.
\end{eqnarray}
The first integral can be precisely given in terms of $-\kappa$ and the second integral due to the large interval can be computed by considering the integrand well approximated by $(1/2)y^{-4}$. Thus we can find the energy given in the form
\begin{eqnarray}\label{En2-reg-2}
 E\simeq2m_q-\frac{U_0\kappa}{\pi}\left(1-\frac16\frac{U_0^3}{\kappa\, U_1^3}\right).
\end{eqnarray}
Now substituting $U_0\simeq 2R^2\kappa/L_0$, i.e., the leading term of Eq.~(\ref{lo2-4}), and $U_1=2\pi m_q$ into Eq.~(\ref{En2-reg-2}) we find
\begin{eqnarray}\label{En2-reg-3}
E\simeq 2m_q-\frac{\alpha^2}{2 L_0}\left(1-\frac{1}{6}\frac{R^6\kappa^2}{\left(\pi L_0 m_q\right)^3}\right),
\end{eqnarray}
where $\frac{\alpha^2}{2}=\frac{2R^2\kappa^2}{\pi}$. Notice that light quarks here stand for a sufficiently small cosmological constant present in $\gamma$ --- See Eq.~(\ref{mass-mq}). This confirms the experimental results pointing the hadronization of light mesons. 

\subsection{Small interquark distance}


In the present case we should consider $U_1/U_0\equiv\sqrt{C}\to1$ \cite{myers}. This is the same condition previously considered in Ref.~\cite{bbq08} do address the confining regime. Again, firstly for $L_0$ we rewrite Eq.~(\ref{lo}) as follows
\begin{eqnarray}\label{eq-L0-small}
\frac{L_0}{2}=\frac{R^2}{U_0}\int^{U_1/U_0\to1}_{1}\frac{dy}{y^2\sqrt{y^4-1}}&\simeq&\frac{R^2}{U_0}\int^{U_1/U_0-1}_{1}\frac{dz}{(1+2z)\sqrt{(1+4z)-1}}\nonumber\\
&\simeq&\frac{R^2}{U_0}\int^{U_1/U_0-1}_{1}\frac{dz}{2\sqrt{z}}\nonumber\\
&\simeq&\frac{R^2}{U_0}\sqrt{\frac{U_1}{U_0}-1}
\end{eqnarray}
where in the r.h.s. we have considered $y=1+z$ and $z\ll1$ in the short interval above. Let us write (\ref{eq-L0-small}) in terms of $U_0$, that is
\begin{eqnarray}\label{eq-L0-small-2}
U_0\simeq U_1\left[1+\frac{U_0^2}{4R^4}L_0^2\right]^{-1}\simeq U_1\left[1-\frac{U_0^2}{4R^4}L_0^2\right]
\end{eqnarray}
Solving iteratively for $U_0$ recalling that $U_0\simeq U_1$ and $U_1=2\pi m_q$ we find
\begin{equation}\label{U0-small}
U_0\simeq 2\pi m_q\left[1-\left(\frac{\pi L_0 m_q}{R^2}\right)^2\right].
\end{equation}

Now for energy we rewrite Eq.~(\ref{En}) as follows
\begin{eqnarray}\label{En2-2}
 E=\frac{ U_0}{\pi}\int_1^{U_1/U_0\to1}\frac{y^2}{\sqrt{y^4-1}}dy&\simeq&\frac{U_0}{\pi}\int_1^{U_1/U_0-1}\frac{1+2z}{\sqrt{(1+4z)-1}}dz\nonumber\\
 &\simeq&\frac{ U_0}{2\pi}\int_1^{U_1/U_0-1}z^{-1/2}dz\nonumber\\
 &\simeq&\frac{ U_0}{\pi}\sqrt{\frac{U_1}{U_0}-1}
\end{eqnarray}
Again, in the r.h.s. we have considered $y=1+z$ and $z\ll1$ in the short interval above. Substituting (\ref{U0-small}) into (\ref{En2-2}) and recalling that $U_1=2\pi m_q$  the energy is now given by the linear form
\begin{equation}
E\simeq\frac{2\pi m_q^2}{R^2}L_0\to E=\sigma_c L_0,\qquad \sigma_c=\frac{\gamma}{\pi\alpha'}
\end{equation}
where in the last step we have used the mass formula (\ref{mass-mq}) --- notice the relationship between $\sigma_c$ and the mass scale $M$ defined in Eq.~({\ref{w-2-M}). It gives the same result obtained in Ref.~\cite{bbq08}. This means a confining regime at sufficiently small interquark distance \cite{myers}.

\section{Conclusions}
\label{conclu}

The cosmological constant that appears involved in the original geometry is attributed to the D7-brane position itself in its process of embedding. As a consequence we find an explicit dependence of the quark mass with the cosmological constant. Analogously, direct computation with the original geometry gives the same interquark potential regimes found in Ref.~\cite{bbq08}. Thus, our results show that we have found a complete equivalence of flavoring with a D7-brane or properly considering a cosmological solution. We also have found an explicit linear dependence of the quark condensate with the quark mass. This seems to be the case for small masses, in the region where the quenching $1/m$ \cite{Fleming:1998cc} still does not take place. So  further investigations related to the quark condensate, for instance, in the present setup should be pursued.


{\acknowledgments} We would like to  thank  L. Barosi, D. Bazeia and A.R. Queiroz for discussions and CNPq, PNPD-CAPES, PROCAD-NF/2009-CAPES for partial financial support.

\end{document}